\documentclass[a4paper,twocolumn,11pt,accepted=2024-02-22]{quantumarticle}
\pdfoutput=1
\usepackage[utf8]{inputenc}
\usepackage[english]{babel}
\usepackage[T1]{fontenc}
\usepackage{amsmath}
\usepackage{hyperref}
\usepackage[numbers]{natbib}
\usepackage[resetlabels]{multibib}
\usepackage{listings}

\usepackage{tikz}
\usepackage{lipsum}
\usepackage{graphicx}
\usepackage{amsmath,amsfonts,amssymb,amsthm}
\usepackage{color}
\usepackage{soul}
\usepackage{subfigure}
\usepackage{physics}
\usepackage{dsfont}
\usepackage{hyperref}
\usepackage{url}
\usepackage{textcomp}
\usepackage{gensymb}
\usepackage{threeparttable}
\usepackage{rotating}
\usepackage{esvect}
\usepackage{booktabs}
\usepackage{marvosym}
\usepackage{mathrsfs}
\usepackage{ytableau}
\usepackage[export]{adjustbox}
\usepackage{mathtools}

\newcites{article}{article references}
\newcites{book}{book references}
\newcites{misc}{misc references}
\newcites{repo}{repository references}
\newcites{web}{website references}
\newcites{other}{Other references}

\newcommand{\GSa}{\includegraphics[trim= 0 2.25cm 0 0, scale=0.3]{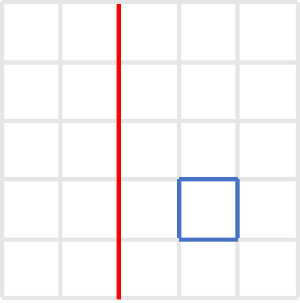}}
\newcommand{\GSb}{\includegraphics[trim= 0 2.25cm 0 0, scale=0.3]{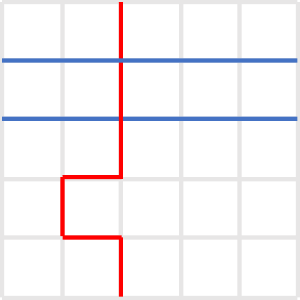}}
\newcommand{\GSc}{\includegraphics[trim= 0 2.25cm 0 0, scale=0.3]{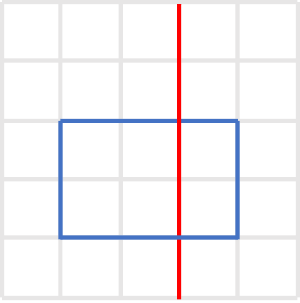}}

\begin{document}

\title{Bicolor loop models and their long range entanglement}

\author{Zhao Zhang}
\affiliation{Department of Physics, University of Oslo, P.O. Box 1048 Blindern, N-0316 Oslo, Norway}
\affiliation{SISSA and INFN, Sezione di Trieste, via Bonomea 265, I-34136, Trieste, Italy}
\orcid{0000-0002-9425-732X}
\email{zhaoz@uio.no}

\maketitle

\begin{abstract}
  Quantum loop models are well studied objects in the context of lattice gauge theories and topological quantum computing. They usually carry long range entanglement that is captured by the topological entanglement entropy. I consider generalization of the toric code model to bicolor loop models and show that the long range entanglement can be reflected in three different ways: a topologically invariant constant, a sub-leading logarithmic correction to the area law, or a modified bond dimension for the area-law term. The Hamiltonians are not exactly solvable for the whole spectra, but admit a tower of area-law exact excited states corresponding to the frustration free superposition of loop configurations with arbitrary pairs of localized vertex defects. The continuity of color along loops imposes kinetic constraints on the model and results in Hilbert space fragmentation, unless plaquette operators involving two neighboring faces are introduced to the Hamiltonian.  
\end{abstract}

\section{Introduction}
\label{sec:Intro}
%
Entanglement is a phenomenon unique to quantum systems as the wave-functions are usually superposition of multiple classical states and can be quantified with the entanglement entropy. In the ground state of a quantum many-body systems, the scaling behavior of entanglement entropy is governed by interaction in the systems and often provides more revealing characterization than the equal time correlation function and the spectral gap. For a gapped system with local Hamiltonian, due to the finite correlation length, entanglement entropy saturates at a finite depth from the boundary of the bipartition and scales linearly with the size of the perimeter. This is referred to as the area law of entanglement entropy and has been proven in one dimension by Hastings~\cite{Hastings_2007}, and in two dimensions for frustration-free spin systems~\cite{AnshuEA21}. However, for a gapless system at quantum critical point, the correlation length diverges and the area law can be violated with an extra logarithmic factor. This has been shown for critical points described by a $(1+1)$-dimensional conformal field theory (CFT)~\cite{Holzhey:1994aa, Calabrese_2009}, and gapless free fermionic systems with a Fermi sea in arbitrary dimension~\cite{GioevEA06}. Unlike gapped Hamiltonians, which dictate the ground state to obey the area law, gapless systems can have entanglement entropy that scales anywhere between satisfying area law and extensively.

Beyond logarithmic violation of the area law of entanglement entropy has been known to arise in two scenarios. The first occurs when degrees of freedom symmetric about the boundary of a bipartition form maximally entangled Bell pairs, due to strong disorder or inhomogeneity of the coupling strength~\cite{Vitagliano_2010, Ramirez_2014, zhang2023entanglement}, which can also be interpreted with a holographic dual Anti-de Sitter spacetime metric~\cite{Rodriguez-Laguna:2017aa, MacCormack:2019aa}. The second is to construct frustration-free parent Hamiltonians with projection operators enforcing superposition of locally different configurations satisfying a certain local geometric constraints, usually with a combinatorics background. In one dimension, examples include the colored Motzkin~\cite{MovassaghShor, Zhang5142} and Fredkin chains~\cite{PhysRevB.94.155140,Salberger:2017aa, Salberger_2017, Zhang_2017}, the entanglement entropy of which has the scaling of the corresponding limit shapes of random Young diagrams. The extensive scaling of entanglement entropy in this case also have a holographic nature in terms of the tensor network description of the ground state~\cite{PhysRevB.100.214430}. In two dimensions, similar quantum models are constructed from classical vertex and dimer or tiling configurations with ring-exchange Glauber dynamics~\cite{zhang2022area, QuantumLozenge}, where entanglement entropy scales according to the limit shape of random surfaces defined by the height function of the U(1) Coulomb gas phase. In fact, with inhomogeneous $q$-deformations of the Hamiltonian, the ground state entanglement can be tuned to any intermediate power law scaling. It is important to note that such area law violations can only happen with the local Hilbert space enlarged by an extra color degree of freedom.

Although sub-area law scaling is trivial for one-dimensional systems, it does lead to interesting physics in two and higher dimensions. In fact, topological entanglement entropy (TEE)~\cite{PhysRevLett.96.110404, PhysRevLett.96.110405} can be viewed as sub-area law correction due to the long range entanglement in models with topological order such as the toric code~\cite{Kitaev:2003aa}. More recently, 2D models with subsystem symmetry have been shown to have spurious topological entanglement entropy lower bounded by the logarithm of the total quantum dimension~\cite{PhysRevB.94.075151,PhysRevLett.122.140506,PhysRevB.100.115112,PhysRevResearch.2.032005,Kim:2023ydi}. As states belonging to the same topological phase with different spurious TEE are connected by constant depth quantum circuit, the spurious TEE cannot scale with the systems size. For a generic area law obeying system, the dependence of entanglement entropy in two dimension on the length of the cut $L$ can be given as
\begin{equation}
	S=\alpha L - \beta \log L -\gamma + o(L^{-1}). 
\end{equation}
In addition to topologically ordered systems, where $\gamma\ne 0$, universal logarithmic sub-leading contributions have been observed in the class of $(2+1)$-dimensional critical points described by conformal field theory (CFT)~\cite{PhysRevLett.97.050404, Casini:2007aa}, which includes examples such as the quantum dimer model~\cite{PhysRevLett.61.2376, PhysRevB.65.024504} and quantum 8-vertex model~\cite{Ardonne:2004aa}. These logarithmic terms are proportional to the Euler-characteristic of the region of the subsystem, or alternatively attributed the corners or curvature of the boundary between subsystems. They can also be computed from the holographic principle using AdS/CFT correspondence~\cite{Hirata:2007aa}, and has been numerically verified for certain lattice models~\cite{PhysRevB.90.235106}. In Ref.~\cite{balasubramanian20232d}, Balasubramanian, Lake and Choi argued a bicolor loop model in its large deformation parameter regime has anomalous TEE that can scale linearly with the perimeter. Motivated by these recent developments, in this manuscript, I study upper bounds on the entanglement entropy of various colored loop models to understand better different kinds of sub-area law scaling due to the long range entanglement in the color degree of freedom.

Loop models have been extensively studied in the context of classical statistical mechanics, especially those with critical points described by CFTs, including both completely packed and dilute loop ones~\cite{Fendley:2006aa}. Entanglement entropy of the frustration-free fully packed loop model has also been recently studied showing a sub-leading logarithmic contribution~\cite{Zhang_2023}. Since Kitaev's seminal work that brought the attention to their relevance in topological quantum computing, loop gas and string-net modals have been heavily investigated in the topological order community, including the $Z_n$ toric code, Kitaev's quantum double~~\cite{Kitaev:2003aa}, double semion and Levin-Wen model~\cite{PhysRevB.71.045110}, and the color code model~\cite{ColorCode, ColorcodeTeo}. However, in none of the existing generalizations to toric code with larger local degrees of freedom, is color a conserved observable along a loop. Partially motivated by the proposal in Ref.~\cite{balasubramanian20232d}, I consider various bicolor loop models, both intersecting and non-intersecting, keeping the continuity of colors of the loops in the ground state. 

Enlarged Hilbert space can usually reveal a deeper reason why things work for a special case with binary degrees of freedom, or lead to much richer physical behaviors that are not expected from a naive generalization. For instance, in one-dimensional spin chains, such generalization can result in integrable excited states in the non-integrable subspace of a partially integrable model that involves scattering between quasiparticles in different colors~\cite{PhysRevB.106.134420}. Another recent example is the generalization of maximally entangled rainbow chain to higher dimensional space with local Hilbert space and Hamiltonians highly similar to the ones used in this manuscript~\cite{zhang2023entanglement}. The model to be discussed in Sec.~\ref{sec:intersec} share the same $S_3$ symmetry as these two models in that the empty uncovered edges are treated on the same footing as those covered in the two colors. In addition, similar to the rainbow chain generalization, each term of the Hamiltonian reproduces the original toric code Hamiltonian when restricted to the subspace involving two of the three components, and acts trivially on the third one. Thus, they are both distinct from models with $Z_3$ parafermion or three copies of Majorana fermions. 

Unlike bicolor dimer models, which has been studied both with update moves of Markov chain Monte Carlo~\cite{Raghavan:1997vi}, and in the context of quantum dimer~\cite{MulticolorDimer}, to the best of my knowledge, the corresponding quantum colored loop model or Glauber dynamics for classical colored loop models has so far been absent.~\footnote{The Balasubramanian--Lake--Choi model~\cite{balasubramanian20232d} may be considered as a first step towards this, but it requires next nearest neighboring interaction involving 12 edges on a square lattice, and since loops in different colors do not intersect, the two colors interact in a way as if the Hamiltonian consists of two copies of unicolor loop Hamiltonians.} Our proposal is therefore the first realization of such a Hamiltonian, constructed in a natural way along the lines of a Stabilizer Hamiltonian. It turns out that due to the intrinsic constraint to maintain the continuity of color in a loop, the model is kinetically constrained. This necessarily leads to drastic differences from the toric code such as Hilbert space fragmentation unless operators acting on more than one plaquettes are added to the Hamiltonian.

The rest of the manuscript is organized as follows. In Sec.~\ref{sec:intersec}, I introduce the Hamiltonian for the intersecting bicolor loop model on the square lattice, and discuss its ground state degeneracy and exact excited states. An upper bound on its bipartite entanglement entropy is evaluated in Sec.~\ref{sec:topoent} to show the topological entanglement entropy in agreement with the quantum dimension. Sec.~\ref{sec:nonintersec} switches gear to define the analogous Hamiltonian for non-intersecting bicolor loops, but on the hexagonal lattice. Upper bound on its entanglement entropy is computed in Sec.~\ref{sec:nontopent}. In Sec.~\ref{sec:longrangeent}, I calculate the upper bounds on entanglement entropy for the Balasubramanian--Lake--Choi model and show it has a non-topological logarithmic sub-leading contribution with the same universal coefficient as the fully packed loop model with one color. Finally, I give a conclusion and discuss the open problems in Sec.~\ref{sec:concl}.

%
\section{The intersecting loop model}
\label{sec:intersec}
%
I start with constructing a quantum Hamiltonian for a bicolor intersecting loop model, based on a natural generalization of the toric code with a three component local degrees of freedom. Like the toric code, the Hamiltonian is frustration free, with solvable ground states being uniform superposition of all intersecting loop configurations with all the edges along a loop in the same color. Unlike the toric code, the topological sectors with different non-contractible loops do not form an irreducible representation of the Wilson loop algebra, as will be shown in Appendix~\ref{sec:Wilson}. Furthermore, it is not clear whether the Hamiltonian is gapped or not, as the plaquette and vertex terms do not commute with each other. Nevertheless some pattern of the spectrum can be deduced from the commutativity between vertex and plaquette operators. 

\subsection{Hamiltonian}

The model is defined on a $L\times L$ square lattice on a torus\footnote{In the following $L$ is assumed to be even. When $L$ is actually odd, the ground state degeneracy will be different.} with local Hilbert space of $\mathbb{C}^3$ living on the edges. Thus, the system consists of $2L^2$ qutrits. The three components correspond to the bond being uncovered, and covered in red or blue color. Define the on-site operators

\begin{equation}
	\begin{aligned}
		X^{(1)}=&\begin{pmatrix}
			0 & 1 & 0\\ 1 & 0 & 0\\ 0 & 0 & 0 
		\end{pmatrix}, 	\quad Z^{(1)}=&\begin{pmatrix}
			1 & 0 & 0\\ 0 & -1 & 0\\ 0 & 0 & 0 
		\end{pmatrix}, \\
		X^{(2)}=&\begin{pmatrix}
			0 & 0 & 1\\ 0 & 0 & 0\\ 1 & 0 & 0 
		\end{pmatrix}, 	\quad Z^{(2)}=&\begin{pmatrix}
			1 & 0 & 0\\ 0 & 0 & 0\\ 0 & 0 &-1 
		\end{pmatrix},	\\
		X^{(3)}=&\begin{pmatrix}
			0 & 0 & 0\\ 0 & 0 & 1\\ 0 & 1 & 0 
		\end{pmatrix},	\quad	Z^{(3)}=&\begin{pmatrix}
			0 & 0 & 0\\ 0 & 1 & 0\\ 0 & 0 & -1 \end{pmatrix},
	\end{aligned}
\end{equation} which are 3-component generalizations to the Pauli $X$, and $Z$ operators with eigenvalues $0,\pm 1$. The Hamiltonian consists of two parts, one as a sum of local operators defined around a vertex $v$
\begin{equation}
	h_v = -\sum_{a=1}^3 A_v^{(a)} + \hat{\Delta}_v,
\end{equation}where
\begin{equation}
	A_v^{(a)}=\prod_{j\in +_v}Z_j^{(a)}, \quad \text{for } a = 1, 2, 3,
\end{equation} and the operator $\hat{\Delta}_v$ checks whether all four edges around vertex $v$ are uncovered or covered in the same color, and returns eigenvalue $1$ if so, but $0$ otherwise. The $A$ operators energetically penalizes discontinuities of colored loops at a vertex, so that only closed loops of a fixed color appear in the ground state, although they can intersect. However, the three vertex configurations in the first row of the left column in Table.~\ref{tab:A} have a lower energy of $-2$ than the otherwise lowest eigenvalue $-1$ of the rest two rows, hence the addition of the $\hat{\Delta}_v$ operators which compensates for the difference. 
\begin{table}
	\begin{center}
		\begin{tabular}{ | c | c | c | } 
			\hline
			$\varepsilon_v=-1$ & $\varepsilon_v=0$ & $\varepsilon_v=1 $ \\ 
			\hline 
			\includegraphics[trim=0 0 0 -0.4cm, height=3.2cm]{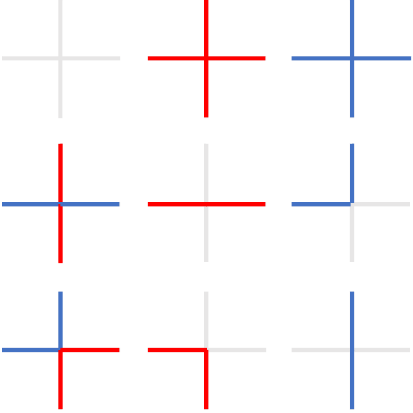} &  \includegraphics[trim=0 0 0 -0.4cm, height=3.2cm]{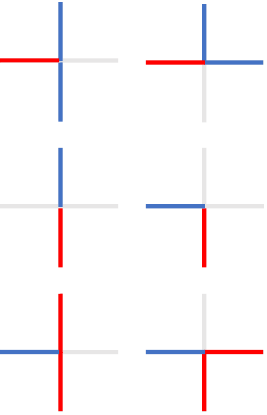} &  \includegraphics[trim=0 0 0 -0.4cm, height=3.2cm]{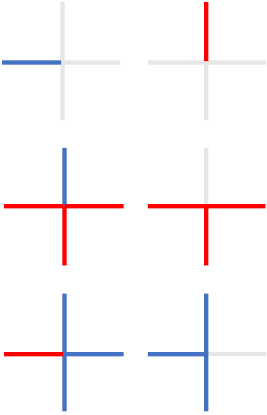} \\ 
			\hline
		\end{tabular}
		\caption{Vertex configurations corresponding to three different eigenvalues of $h_v$, up to rotations and reflections.\label{tab:A}}
	\end{center}
\end{table}

\begin{figure}
	\centering
	\subfigure[]{\label{fig:B2}\includegraphics[width=0.6\linewidth]{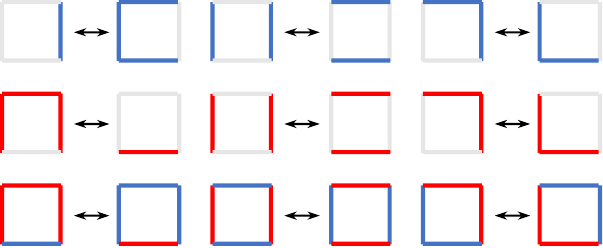} }
	
	\subfigure[]{\label{fig:B1}\includegraphics[width=0.17\linewidth]{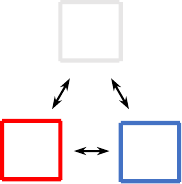} }
	\hspace{1cm}
	\subfigure[]{\label{fig:B3}\includegraphics[width=0.26\linewidth]{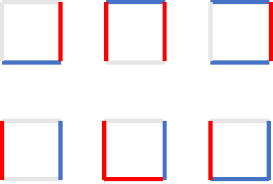} }
	\caption{Face configurations (up to rotations and reflections) that appear in the eigenstates of $h_f$ with minimal eigenvalue, as superpositions of (a) symmetrized pairs; (b) symmetrized triplets; and (c) singlets.}
	\label{fig:B}
\end{figure}

The other part of the Hamiltonian is defined as a sum of local off-diagonal operators around a face $f$
\begin{equation}
	h_f = -\sum_{a=1}^3 B_f^{(a)} - \hat{N}_f,
\end{equation} with
\begin{equation}
	B_f^{(a)}=\prod_{j\in \square_f}X_j^{(a)}, \quad \text{for } a = 1,2,3,
\end{equation} and the operator $\hat{N}_f$ counts the number of different configurations around face $f$ and returns eigenvalues $1,2$ or $3$ accordingly. The explicit form of $\hat{N}_f$ is not particularly illuminating, but it can be easily written down in the diagonal basis of the color indices. The $B$ operators favor superposition of locally different loops by introducing the off-diagonal terms that relates them in pairs of two or three, as shown in Fig.~\ref{fig:B}. However, faces involving edges in three different colors (Fig.~\ref{fig:B3}) are annihilated by the $B$ operators, and those involving a single color or uncovered (Fig.~\ref{fig:B1}) have a different eigenvalue than those involving two (Fig.~\ref{fig:B2}). Therefore the $\hat{N}_f$ operator needs to be included to make the plaquette Hamiltonian frustration free.

The eigenstates of the Hamiltonian 
\begin{equation}
	H=\sum_v h_v +\sum_f h_f
	\label{eq:interH}
\end{equation} 
are superpositions of bicolor closed loop configurations related by sequential flipping of local plaquette configurations, and have degenerate energy $-4L^2$, as $h_v$ has lowest eigenvalue $-1$ and the lowest eigenvalue of $h_f$ is $-3$.

\subsection{Ground state degeneracy}

\begin{figure}[t!bh]
	\centering
	\includegraphics[trim=0 0 0 -8mm,width=\linewidth]{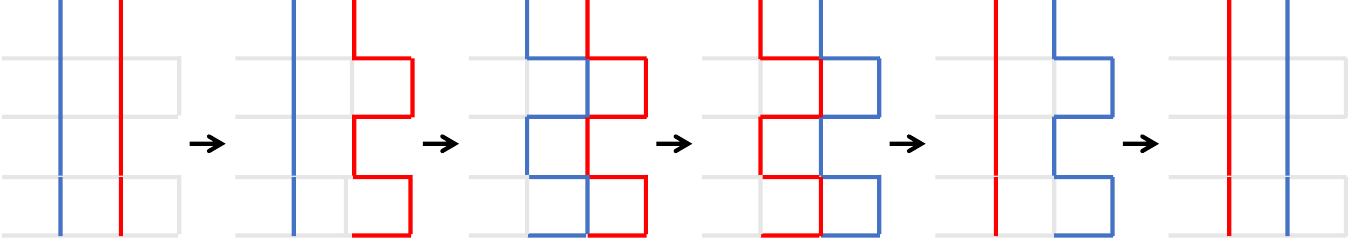}
	\caption{The sequence moves applying $B$ operators that swaps two neighboring non-contractible loops.}
	\label{fig:noncontract}
\end{figure}

As in the toric code model, an immediate consequence of the torus topology is that non-contractible loops in both direction of the lattice separate the ground state subspace into disconnected topological sectors, as the off-diagonal $B_f$ operators are not able to relate loop configurations differing by non-contractible loops. They are, however, able to deform non-contractible loops, and create or annihilate them in pairs when they are adjacent. The possibility of swapping non-contractible loops around each other is not so obvious given the absence of kinetic terms involving face configurations in Fig.~\ref{fig:B3}. However, it can be done by the sequence of moves depicted in Fig.~\ref{fig:noncontract}. This gives a 4-fold degeneracy in either direction, characterized by the color of the non-contractible loops that appear an odd number of times, which must have an even number of either 0 or 2 in total
\begin{equation*}
	\{\ket{\emptyset}, \ket{\mathrm{r}},\ket{\mathrm{b}},\ket{\mathrm{rb}} \},
\end{equation*}
where in $\ket{\mathrm{r}}$, and $\ket{\mathrm{b}}$, the empty non-contractible loop also appears an odd number of times. Taking both directions of the torus into account, this gives a ground state degeneracy of $4^2=16$.

\begin{figure}[t!bh]
	\centering
	\subfigure[]{\label{fig:noncontraexception}\includegraphics[width=0.4\linewidth]{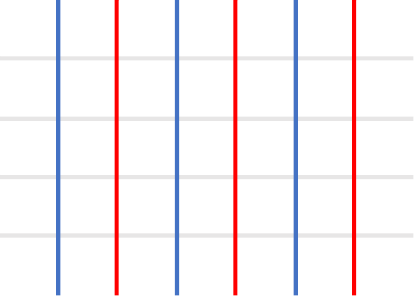} }
	\subfigure[]{\label{fig:fragment}\includegraphics[width=0.4\linewidth]{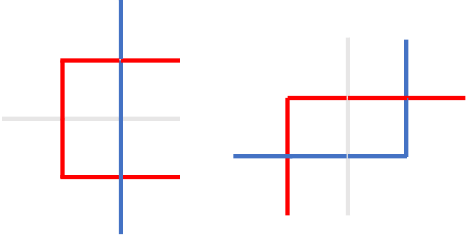} }
	\caption{The frozen local configurations without the addition of $C_{<f,f'>}$ operators in the Hamiltonian.}
	\label{fig:counterexample}
\end{figure}

Clearly, there is a catch with this argument. That is the possibility of such sequences of moves depends on the configuration outside the neighborhood. In fact, there is an obvious example of alternating non-contractible loops occupying all columns, as shown in Fig.~\ref{fig:noncontraexception}, where non of the $B_f^{(a)}$ operators can act non-trivially. Hence it is an isolated state in the Hilbert space, forming a degenerate ground state as an unentangled product state. In addition, the local configurations in Fig.~\ref{fig:fragment} are also frozen, in that the $B_f^{(3)}$ operators do not help untangle the red and blue loops when they are pierced through by an empty loop.

Fortunately, both types of fragmentation disappear with the addition of off-diagonal operators acting on two neighboring faces
\begin{equation}
	C_{<f,f'>}^{(a)}=\prod_{j\in (f\cup f')\setminus (f \cap f')}X_j^{(a)}, \quad \text{for } a=1,2,3.
\end{equation}
As we will show in Appendix~\ref{sec:ergodicity}, the Hamiltonian
\begin{equation}
	H_{\mathrm{int}}=\sum_v h_v +\sum_f h_f + \sum_{<f,f'>}h_{<f,f'>}
	\label{eq:totalH}
\end{equation} is frustration free and has a unique ground state within each topological sector, if
\begin{equation}
	h_{<f,f'>}=-\sum_{a=1}^3C_{<f,f'>}^{(a)}-\hat{N}_{<f,f'>},
\end{equation} and $\hat{N}_{<f,f'>}$ is a diagonal operator that returns the number of different states around the 6 edges in $(f\cup f')\setminus (f \cap f')$ for the same consideration as in $h_f$. The ground state energy of Hamiltonian~\eqref{eq:totalH} is $-10L^2$, as there are two types of double plaquette operators, acting on two horizontally or vertically adjacent plaquettes, and each of them has a lowest eigenvalue of $-3$. The ground state is 16-fold degenerate, and an example in the $\ket{\emptyset_x\mathrm{r}_y}$ sector can be expressed as
\begin{equation}
	\begin{split}
		\ket{\mathrm{GS}_{\emptyset_x\mathrm{r}_y}}=\frac{1}{\sqrt{\mathcal{N}_{\emptyset_x\mathrm{r}_y}}}\left(\ket{\GSa}+\ket{\GSb}\right.\\ \left.+\ket{\GSc}+\cdots\right),
	\end{split}
\end{equation}
where the normalization constant $\mathcal{N}_{\emptyset_x\mathrm{r}_y}$ is the number of intersecting loop configurations with a net non-contractible red loop in the $y$ direction.

\subsection{Exact excited states}

Despite the commutation relations 
\begin{equation}
	[A_v^{(a)}, B_f^{(a)}]=0,  \qquad \forall v,f,
\end{equation}
following the stabilizer code in the respective subspace of $a=1,2,3$, it is not surprising that the plaquette operators do not commute among themselves
\begin{equation}	
	[B_f^{(a)}, B_f^{(b)}]\ne0,
\end{equation}
for $a\ne b$. However, what is not so obvious is that
\begin{equation}
	\begin{aligned}
		[A_v^{(a)}, B_f^{(b)}]&=0,\\
		[A_v^{(a)}, C_{<f,f'>}^{(b)}]&=0,
	\end{aligned}
\end{equation}
for $a\ne b$ as well. This is because the plaquette operators either annihilate a state regardless of acting before or after the vertex operators, or simply swaps the states of two of the legs around a vertex, leaving the eigenvalue of the $A^{(a)}$ operators unchanged. Separating the vertex and plaquette parts of the Hamiltonian \eqref{eq:totalH} as
\begin{equation}
	\begin{aligned}
		H_v=&\sum_v h_v,\\
		H_f=&\sum_f h_f + \sum_{<f,f'>} h_{<f,f'>},
	\end{aligned}
\end{equation}
the above can be summarized as 
\begin{equation}
	[H_v,H_f]=0.
\end{equation}
Unlike the toric code, $H_f$ is not exactly solvable except for its lowest energy eigenstate. 

\begin{figure}[t!bh]
	\centering
	\includegraphics[width=0.8\linewidth]{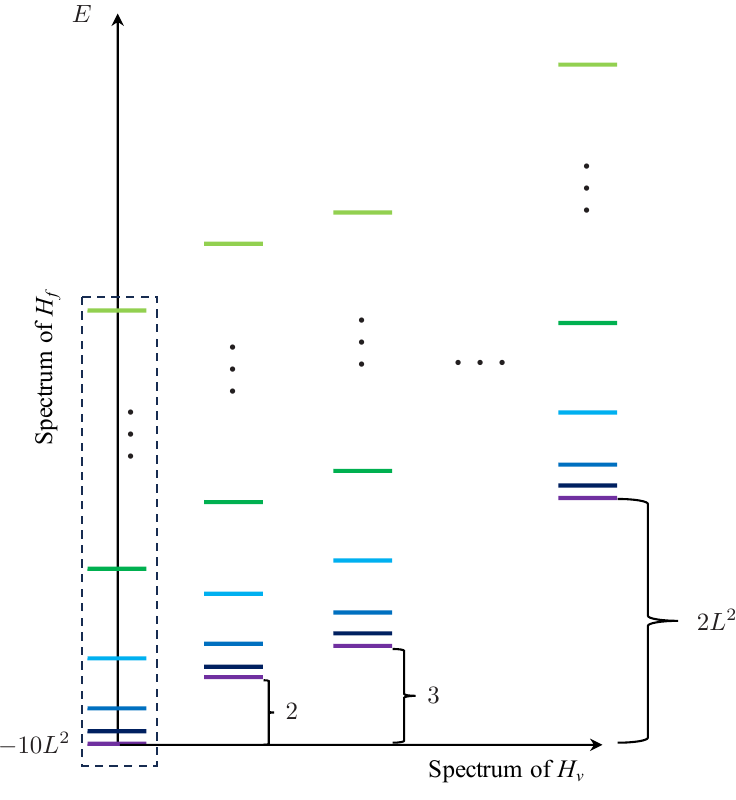}
	\caption{Structure of the spectrum of $H_{\mathrm{int}}$: for each eigenstate of the chaotic spectrum of $H_f$, there is a tower of excited state (marked by the same color) corresponding to the excited states of $H_v$ with pairs of vertex defects. }
	\label{fig:spectrum}
\end{figure}

The Hamiltonian \eqref{eq:totalH} can be viewed as a special case of Shiraishi and Mori's embedding construction of Eigenstate Thermalization Hypothesis (ETH) violating Hamiltonians~\cite{PhysRevLett.119.030601}. But there is more structure to the spectrum. As shown in Fig.~\ref{fig:spectrum}, the spectrum of $H_{\mathrm{int}}$ decomposes into copies of the spectrum of $H_f$. Each copy is shifted by an integer energy according to the eigenvalues of $H_v$. Notice that as there are vertex defects of different energy in Table~\ref{tab:A}, although defects are created in pairs, the pair does not have to be of the same type, which explains the odd integer valued energies in the spectrum.

It is important to point out that although towers of excited states equal distant in energy exist for all eigenstates of $H_f$, only the one generated by introducing vertex defects to the ground state obeys the area law of entanglement entropy, while the other towers will generically have extensive entanglement entropy. Such a structure in spectrum is ubiquitous in two dimensional kinetically constrained models, including the quantum dimer models and quantum fully-packed loop model~\cite{Zhang_2023}. In these models, there are usually diagonal vertex operators enforcing local constraints in the ground state, so that the ground state is a superposition of all classical  configurations satisfying these constraints. Violating those constraints by introducing vertex defects therefore results in a tower of area-law obeying excited states equal distant in energy.

\section{Topological entanglement entropy}
\label{sec:topoent}

Although analytically computing the number of intersecting bulk loop configurations is hard for a given boundary condition, it is illuminating to evaluate an upper bound on the bipartite entanglement entropy from the maximal number of allowed boundary configurations between two subsystems, under the constraint of closed loops. The Schmidt decomposition for a division into subsystems $\mathcal{A}$ and $\mathcal{B}$ along a cut of length $2l$ can be written as
\begin{equation}
	\ket{\mathrm{GS}}=\sum_{\sigma\in\{\phi, \mathrm{r,b}\}^{\otimes {2l}}}\sqrt{p_\sigma}\ket{\mathrm{GS}_\mathcal{A}(\sigma)}\otimes\ket{\mathrm{GS}_\mathcal{B}(\sigma)},
\end{equation}
where $\sigma$ denotes the string of configurations along the cut of length $2l$ and $\phi$ denotes the empty state of an edge. The precise location of the cut is picked close to the edge it is parallel to, so that those it intersect are considered part of subsystem $\mathcal{A}$.

\begin{figure}[t!bh]
	\centering
	\includegraphics[width=0.5\linewidth]{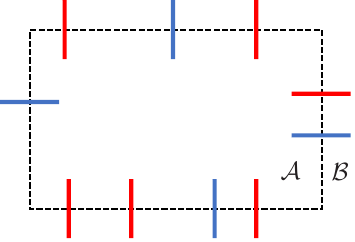}
	\caption{Configurations on the cut between subsystem $\mathcal{A}$ and $\mathcal{B}$ where colored edges don't have to come in pairs along the cut as loops can intersect in the ground state superposition.}
	\label{fig:intersect}
\end{figure}
The Schmidt coefficients $p_\sigma$'s are only nonzero when the configuration $\sigma$ satisfy the constraint that the number of red and blue edges crossing are both even, which is the necessary and sufficient condition for it be possible to form closed loops. Moreover, the maximal entanglement allowed in this case happens when all the non-vanishing Schmidt coefficients equal to the inverse of the total number of allowed configurations at the interface $N_i$. This number can be evaluated from the average of the trinomial expansions
\begin{equation}
	(1+s_\mathrm{r}+s_\mathrm{b})^{2l}=\sum_{i+j=0}^{2l}\binom{2l}{i\quad j\quad 2l-i-j} s_\mathrm{r}^i s_\mathrm{b}^j,
\end{equation}	
for all 4 choices of $s_\mathrm{r},s_\mathrm{b}=\pm 1$, which gives the sum of trinomial coefficient with all three factors appearing in even powers
\begin{equation}
	N_\mathrm{i}=\sum_{i,j=0\mod{2}}\binom{2l}{i\quad j\quad 2l-i-j}=\frac{3^{2l}+3}{4}.
\end{equation} Taking $p_\sigma=\frac{1}{N_\mathrm{i}}$ for all allowed configurations, we have the upper on entanglement entropy
\begin{equation}
	S_\mathrm{i}=-\sum p_\sigma \log p_\sigma = 2l\log 3 -\log4 +\epsilon,
\end{equation} where $\epsilon$ vanishes in the limit $l\to \infty$. Likewise, the upper bound on entanglement entropy in the other topological sectors can be computed to reveal the same constant piece. The topological term agrees with the quantum dimension of $\mathcal{D}=4$ for an Abelian topological order with 16-fold ground state degeneracy.

\section{The non-intersecting loop model}
\label{sec:nonintersec}

A bicolor loop model can be defined on the hexagonal lattice to prevent the loops from intersecting each other. The vertex operator can be written as 
\begin{equation}
	h'_v=-\sum_{a=1}^2 {A'}_v^{(a)} +\hat{\Delta}'_v,
\end{equation}	
where the ${A'}_v^{(a)}$ operators are now defined as product of the $Z^{(a)}$ operators on the 3 edges attached to vertex $v$ instead of 4, and likewise $\hat{\Delta}'_v$ checks if all three legs of $v$ are in the same color. The new vertex operator has the eigenvalues listed in Table~\ref{tab:Ap} for different vertex configurations.

\begin{table}
	\begin{center}
		\begin{tabular}{ | c | c |} 
			\hline
			$\varepsilon_v=-1$ & \includegraphics[trim=0 0 0 -0.4cm, height=1cm]{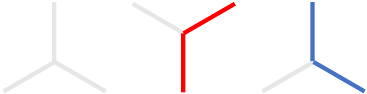}\\ 
			\hline 
			$\varepsilon_v=0$ &  \includegraphics[trim=0 0 0 -0.4cm, height=1cm]{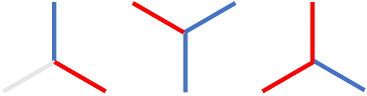} \\
			\hline
			$\varepsilon_v=1$ &    \includegraphics[trim=0 0 0 -0.4cm, height=1cm]{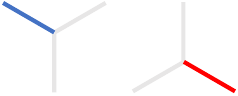} \\ 
			\hline
			$\varepsilon_v=2$ &    \includegraphics[trim=0 0 0 -0.4cm, height=1cm]{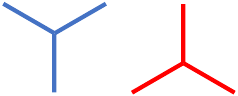} \\ 
			\hline
		\end{tabular}
		\caption{Vertex configurations corresponding to 4 different eigenvalues of $h'_v$, up to rotations and reflections.\label{tab:Ap}}
	\end{center}
\end{table}

\begin{figure}[t!bh]
	\centering
	\includegraphics[width=0.9\linewidth]{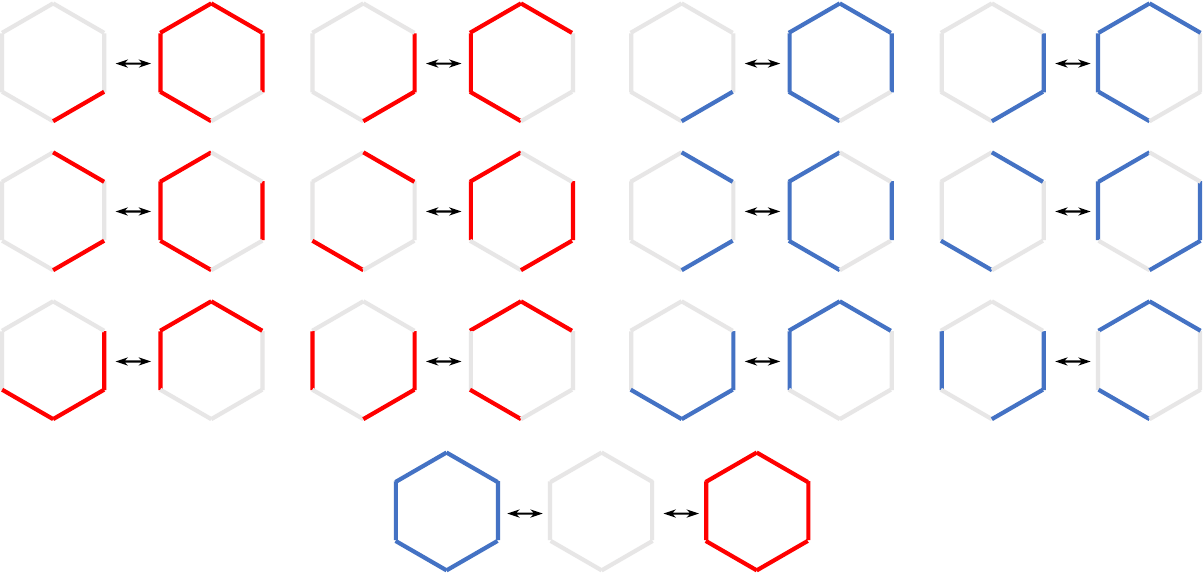} 
	\caption{Non-vanishing off-diagonal entries of the $h'_f$ operators, up to rotations and reflections.}
	\label{fig:Bp}
\end{figure}

The plaquette operators are defined as products of the six $X^{(a)}$ operators around hexagonal faces
\begin{equation}
	h'_f=-\sum_{a=1}^2 {B'}_f^{(a)} + \hat{U}^{(0)}_f,
\end{equation}
where the diagonal $\hat{U}^{(0)}_f$ operator assigns an energy $1$ to faces with all 6 edges uncovered, and $0$ otherwise. The non-vanishing off-diagonal entries are shown in Fig.~\ref{fig:Bp}.

\subsection{Ground state degeneracy}

\begin{figure}[t!bh]
	\centering
	\includegraphics[width=0.5\linewidth]{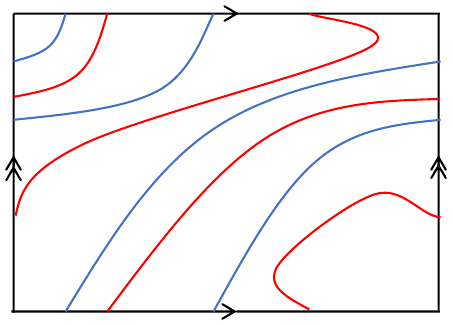} 
	\caption{Non-contractible loops in both directions in the non-intersecting bicolor loop model, where periodic boundary condition and torus geometry are indicated by the arrow and double-arrows.}
	\label{fig:xynoncontra}
\end{figure}

Unlike the intersecting bicolor loop model, the Hamiltonian connects all self-avoiding loop configurations in a certain topological sector, and the only degeneracy comes from non-contractible loops. However, the topological degeneracy now is extensive, since non-contractible cycles in different color cannot pass through each other. The admissible non-contractible loop combinations are
\begin{equation}
	\begin{split}
		\{&\ket{\emptyset_x,\emptyset_y},\\ &\ket{\emptyset_x}\otimes\{\ket{\mathrm{r}_y},\ket{\mathrm{b}_y},\ket{\mathrm{r}\mathrm{b}_y},\ket{\mathrm{r}\mathrm{b}\mathrm{r}\mathrm{b}_y},...\},\\
		&\{\ket{\mathrm{r}_x},\ket{\mathrm{b}_x},\ket{\mathrm{r}\mathrm{b}_x},\ket{\mathrm{r}\mathrm{b}\mathrm{r}\mathrm{b}_x},...,\}\otimes\ket{\emptyset_y},\\
		&\{\ket{\mathrm{r}_x,\mathrm{r}_x},\ket{\mathrm{b}_x,\mathrm{b}_x},\ket{\mathrm{r}\mathrm{b}_x,\mathrm{r}\mathrm{b}_y},\ket{\mathrm{r}\mathrm{b}\mathrm{r}\mathrm{b}_x,\mathrm{r}\mathrm{b}\mathrm{r}\mathrm{b}_y},...\}\},
	\end{split}
\end{equation}
where the last line corresponds to non-contractible loop configurations like those depicted in Fig.~\ref{fig:xynoncontra}.

The uniqueness of the ground state in each topological sector is trivial to prove in the non-intersecting case. First, starting from loops not enclosing smaller ones, all contractible loops can be shrunk to one hexagonal face by the isotopy moves in the upper panel of Fig.~\ref{fig:Bp}, and subsequently removed by the move in the lower right panel. Once there is only non-contractible loops left, topologically invariant configurations can be brought to each other with isotopy moves.

\section{Reduced bond dimension}
\label{sec:nontopent}

Following the same strategy as in Sec.~\ref{sec:topoent}, I compute an upper bound on the bipartite entanglement in the non-intersecting bicolor loop model to show that the coefficient of the area law scaling is smaller than the bond dimension. Since the loops that intersect with the boundary between two subsystems are non-intersecting, the edge configuration along the boundary not only have to have even number for each color in total, but it also cannot have an odd number of edges in the same color sandwiched between edges of the other color. Such configurations can be enumerated with the help of a transfer matrix of the stochastic process among the five possible states $\{\ket{\phi}, \ket{\mathrm{r}},\ket{\mathrm{b}},\ket{\mathrm{rb}},\ket{\mathrm{br}}\}$, where the states encodes the history with the leftover edges in each color yet to be paired up with. The transfer matrix in this basis is given as
\begin{equation}
	T=\begin{pmatrix}
		1 & 1 & 1 & 0 & 0\\
		1 & 1 & 0 & 0 & 1\\
		1 & 0 & 1 & 1 & 0\\
		0 & 0 & 1 & 1 & 0\\
		0 & 1 & 0 & 0 & 1
	\end{pmatrix},
\end{equation}
which has eigenvalues $\{\sqrt{3}+1,2,1,0,1-\sqrt{3}\}$. The total number of allowed configuration for a boundary of length $2l$ is thus $N_\mathrm{n}=(\sqrt{3}+1)^{2l}+2^{2l}+1+(1-\sqrt{3})^{2l})$. The upper bound on entanglement entropy given by an equal distribution of Schmidt coefficients among these possible boundary configurations is thus 
\begin{equation}
	S_\mathrm{n}=2l\log (1+\sqrt{3}) +\epsilon.
\end{equation} 

\section{Logarithmic subleading correction}
\label{sec:longrangeent}

So far, I have constructed two examples of colored loop models, and obtained upper bounds on the respective entanglement entropy of their ground states. The intersecting bicolor loop model has a topological entanglement entropy in agreement with the topological ground state degeneracy. The non-intersecting bicolor loop model, however, does not have such a topological term in the entanglement entropy, but the bond dimension of the area law term is smaller than the dimension of the local Hilbert space. In this section, I obtain a third type of upper bound on entanglement entropy with a sub-leading logarithmic correction to the area law, from a different non-intersecting bicolor loop model introduced in Ref.~\cite{balasubramanian20232d}. The anomalous scaling can be attributed to the emergent height function due to the absence of intersection and well-defined interior and exterior of loops marked by the perpendicular arrows attached to edges, which can be captured by a scalar field theory, and has been studied in the context of entanglement entropy of 2D conformal critical points~\cite{PhysRevLett.97.050404} and free fields~\cite{Casini:2007aa,Casini:2007aa}. I start by reviewing the similar upper bound on the (unicolor) fully-packed loop model~\cite{Zhang_2023}.

\subsection{The fully-packed loop model}

The upper bound on entanglement entropy was obtained for the fully packed loop model in Ref.~\cite{Zhang_2023}. Due to the boundary condition in that model, the total number of boundary configuration along the cut of length $2l$ is given as $N_{\mathrm{FPL}}=\binom{2l}{l}$, which in the large $l$ limit can be approximated by the Sterling formula as 
\begin{equation}
	N_{\mathrm{FPL}}\approx \frac{2^{2l}}{\sqrt{\pi l}}.
\end{equation}Consequently, an upper bound on the entanglement entropy is \begin{equation}
	S_{\mathrm{FPL}}=2l \log 2-\frac{1}{2}\log (2l)-\frac{1}{2}\log\frac{\pi}{2}, 
	\label{eq:FPLscaling}
\end{equation} assuming equal distribution of Schmidt coefficients among the space of boundary configurations.

\subsection{The Balasubramanian--Lake--Choi model}
\label{sec:Choimodel}

The Balasubramanian--Lake--Choi model~\cite{balasubramanian20232d} is also a model of non-intersecting bicolor loops, which was argued heuristically in their paper to give rise to an anomalous topological entanglement entropy, by which the authors mean that the topological entanglement entropy computed from the Kitaev-Preskill or Levin-Wen scheme is sensitive to deformations of the boundary. Their model differs from the model introduced in Sec.~\ref{sec:nonintersec} in two ways. Firstly, the constraint of no intersection in their model is enforced by involving next nearest neighbor interactions, in the sense that the off-diagonal terms of the Hamiltonian acting on one face of the lattice is conditioned by the state the other eight legs on its four vertices, whereas the model in Sec.~\ref{sec:nonintersec} avoided that with the intrinsic property of hexagonal lattice. Secondly, and more importantly, their model does not allow surgery moves, but only isotopy and creation/annihilation. This can be most explicitly seen from the fact that their colored edges have perpendicular arrows attached to them which always points towards inside the loop. This results in a different enumeration of the number of configurations along the cut and a qualitatively different scaling behavior of entanglement, which has an upper bound with a sub-area-law logarithmic correction.

\begin{figure}[t!bh]
	\centering
	\includegraphics[width=0.9\linewidth]{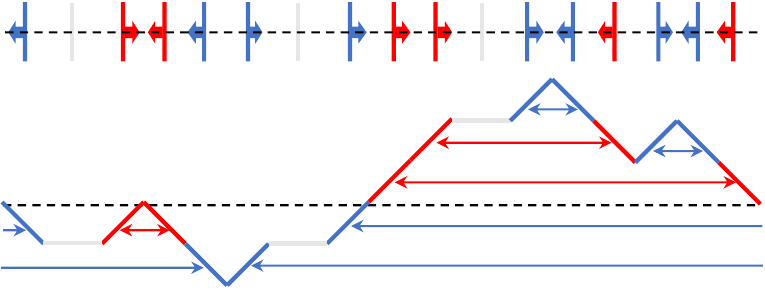} 
	\caption{Mapping from the loop configuration along a cut between subsystems to a random walk returning to origin.}
	\label{fig:mapping}
\end{figure}

The enumeration of such boundary configurations can be mapped to the one-dimensional combinatorial problem of random walks with two types of up and down moves, and returns to origin after $2l$ steps. Picking a particular starting point on the boundary, and counting clockwise, an empty edge is mapped to a flat move, and the two colors of loops depending on the perpendicular arrow attached to the edge differentiating inside and outside of the loop are mapped to up and down moves respectively for both colors. This map is bijective, as can be seen from Fig.~\ref{fig:mapping}. To get a more illuminating result, I enumerate such random walk configurations with $s$ number of coloring of loops, which is given by
\begin{equation}
	N_{\mathrm{BLC}}=\sum_{n=0}^l\binom{2l}{n \quad n \quad 2l-2n}s^n.
\end{equation}

Its asymptotic behavior can be extracted with saddle point approximation.\footnote{Incidentally, this expression is the hypergeometric function ${}_2F_1(\frac{1}{2}-l,-l,1,8)$.} In the large $l$ limit, the summand maximizes at 
$n \approx \sigma l$, with $\sigma = \frac{4s-2\sqrt{s}}{4s-1}$.

Using the Sterling formula, the summand can be approximated as
\begin{equation}
	\begin{aligned}
		&\binom{2l}{\sigma l+x\quad \sigma l+x\quad 2(1-\sigma)l -2x}s^n\\
		\approx&\frac{1}{2\pi(\sigma l +x)}\sqrt{\frac{l}{(1-\sigma)l-x}}(1-\sigma)^{-2l}\times\\
		& \Big(\frac{l}{ l+\sigma^{-1}x}\Big)^{2(\sigma l+x)}\Big(\frac{l}{l-(1-\sigma)^{-1}x}\Big)^{2[(1-\sigma)l-x]},
	\end{aligned}
	\label{eq:sterling}
\end{equation}
where I have used that 
\begin{equation}
	\frac{\sigma}{2}=\frac{2s-\sqrt{s}}{4s-1}\equiv\sqrt{\sigma}(1-\sigma).
\end{equation} The last line of \eqref{eq:sterling} can be approximated as
\begin{widetext}
\begin{equation}
	\begin{aligned}
		&\exp{-2(\sigma l+x)\log(1+\frac{x}{\sigma l})}\times\exp{-2\big((1-\sigma)l-x\big)\log(1-\frac{x}{(1-\sigma)l})}\\
		\approx&\exp{-2(\sigma l+x)\Big(\frac{x}{\sigma l}-\frac{x^2}{2\sigma^2 l^2}\Big)}\times\exp{-2\big((1-\sigma)l-x\big)\Big(-\frac{x}{(1-\sigma)l}-\frac{x^2}{2(1-\sigma)^2l^2}\Big)}\\
		\approx&\exp{-\frac{x^2}{\sigma l}-\frac{x^2}{(1-\sigma)l}}.
	\end{aligned}
\end{equation}
\end{widetext}
Since the summand is suppressed by a Gaussian distribution around a point away from the boundary of the summation, we can approximate it with an integral from $-\infty$ to $\infty$ and omit the $x$'s in the prefactors of \eqref{eq:sterling}
\begin{widetext}
\begin{equation}
	\begin{aligned}
		N_{\mathrm{BLC}}\approx& \frac{1}{4\pi \sqrt{s} l}(\frac{2\sqrt{s}}{\sigma})^{2l+\frac{3}{2}} \int_{-\infty}^{\infty}dx \exp{-\frac{\big((2\sqrt{s}+1)x\big)^2}{2\sqrt{s}l}}\\
		\approx& \frac{(2\sqrt{s}+1)^{2l+\frac{1}{2}}}{2\sqrt{2\pi l}}.
	\end{aligned}
\end{equation}
\end{widetext}
Assuming a uniform distribution of Schmidt coefficients, an upper bound of entanglement entropy can be obtained as
\begin{multline}
	S_{\mathrm{BLC}}=2l\log(2\sqrt{s}+1)-\frac{1}{2}\log (2l)\\ -\frac{1}{2}\log\frac{4\pi}{2\sqrt{s}+1},
\end{multline}	
which has the anomalous logarithmic correction that is \emph{not} topological in nature, in the sense that the TEE computed in the Kitaev-Preskill or Levin-Wen scheme is not invariant to deformations of the boundaries. The coefficient of the logarithmic correction is the same as in \eqref{eq:FPLscaling}, which can be attributed to the emergent height function due to the loops serving as contour lines as in a topographic map. It should be remarked that the calculation in this section differs from the argument in Ref.~\cite{balasubramanian20232d} in that while the authors there focused on the highly deformed regime of their parameter space, in which nested loops are favored in the ground state superposition, the calculation here is done for the un-deformed point in the parameter space, for which the upper bound is supposedly tightest. It is expected, however, that regardless of the deformation parameter, the actual entanglement entropy does not follow any scaling law for the subleading contributions in this model.

\section{Discussions}
\label{sec:concl}

In this work, I constructed Hamiltonians for two bicolor loop models, one intersecting on the square lattice, and the other non-intersecting on the honeycomb lattice. Their degenerate ground states form different topological sectors, and topological entanglement entropy only appears for the intersecting model, while the ground state entanglement entropy of the non-intersecting model is upper bounded by an area law scaling with a smaller bond dimension than the local degrees of freedom. I further obtained an upper bound on the entanglement entropy of the Balasubramanian--Lake--Choi bicolor loop model, where a height function along the bipartition boundary emerges from the intrinsic obstruction of cutting and gluing loops. The subleading logarithmic contribution to the entanglement entropy is shown to have the same coefficient as the fully packed loop model, the height function of which in the scaling limit is known to be described by a Gaussian free field theory. Yet, I argue that the appearance of logarithmic subleading term is different from the corner contribution of geometric curvature of the manifold in (2+1)D CFT, which was known before~\cite{PhysRevLett.97.050404, Casini:2007aa}.

Enlarging the local Hilbert space with the color degree of freedom poses new questions to the well understood loop gas model of toric code. First, unlike the toric code our Hamiltonian consists of non-commuting plaquette operators and are therefore not exactly solvable for the whole spectrum. Although the vertex operators do commute with the plaquette part of the Hamiltonian, the spectrum of the latter is not understood except for its frustration free ground state. In particular, it is not clear whether there will be a spectral gap between the ground state and the lowest energy excited state in the thermodynamic limit. Given that the bicolor loop models bear enough resemblance to be viewed as a higher-dimensional cousin of the pair-flip model~\cite{caha2018pair}, one may try to apply the techniques employed there to obtain upper and lower bounds on the spectral gap. One can also try to compute the loop correlation function either numerically or analytically, which is expected to decay algebraically, as opposed to the exponential decal of spin-spin correlations.

Second, without the introduction of plaquette operators acting on two neighboring faces, the kinetic constraints imposed by the continuity of color in loops would result in Hilbert space fragmentation. In Appendix~\ref{sec:ergodicity}, it is proven that double plaquette operators alone suffices for ergodicity of bicolor intersecting loops. It would be interesting to explore how that changes when more colors are involved, in particular whether operators with finite support could guarantee ergodicity and if so, how the support need to grow with the number of colors. Another direction to pursue further is to find either topological or some other type of invariants that characterize the resulting Krylov subspaces without involving operators acting on multiple plaquettes. The situation here is similar to Brunnian links, as the intersection between any two types of loop is trivial in the absence of the third type, but non-trivial when all three are involved. Yet another potentially relevant topological object is the three loop braiding in three dimension~\cite{PhysRevLett.113.080403}. As for the non-intersecting loop model, although the ground state manifold only consists of different topological sectors, the number of degeneracy that grows with system size might also lead to interesting topological field theory description.

Third, the tower of exact excited states corresponding to the uniform superposition of loop configurations with the same vertex defect might lead to the persistent oscillation of certain initial states in time evolution. To determine whether this is indeed the case, one need to study whether the tower of eigenstates are related to some spectrum generating algebra~\cite{PhysRevB.101.195131} and examine if the operators involved are pseudo-local in the sense defined by Ref.~\cite{Doyon:2017aa} and \cite{buvca2023unified}. 

Fourth, it would be interesting to pursue interactions at the atomic level that might allow the colored loop Hamiltonian to emerge. One potential path was discussed in Ref.~\cite{stahl2023topologically}, from the duality between the quad-flip model and generalized PXP models. Alternatively, one might be able to follow the Kitaev's paradigm to obtain a qutrit toric code as an effective Hamiltonian of certain phases of a generalized honeycomb model~\cite{KITAEV20062}.

Finally, I emphasize that the upper bounds on entanglement entropy for the ground states of various models, although believed tight, can differ from the actual entanglement entropy. Exact enumeration of bulk configurations corresponding to the same boundary configuration either analytically or numerically might reveal more detail about the universality of the sub-leading contribution and the nature of the logarithmic correction.

\acknowledgments
	I thank Shankar Balasubramanian, Berislav Bu\v{c}a, Libor Caha, Soonwon Choi, Paul Fendley, Israel Klich, Ethan Lake, Henrik R{\o}ising, Maksym Serbyn, Germ\'{a}n Sierra, Olav Syljuåsen, Jeffrey Teo, and Erik Tonni for fruitful discussions.

\bibliographystyle{quantum}
\bibliography{bicolorloop}

\appendix

\section{Ergodicity and frustration freeness}
\label{sec:ergodicity}

In this appendix, I first prove that the Hamiltonian \eqref{eq:totalH} is frustration free, showing the existence of solvable ground states. Then I establish the ergodicity of the off-diagonal single and double plaquette operators within each topological sector, concluding that the ground state is unique within each sector, giving a 16-fold topological degeneracy.

\subsection{Frustration freeness}

It is clear that there is no competition between the vertex operators and the plaquette or double plaquette operators, as one can always find the correct superposition within the subspace satisfying the vertex constraints. To further show that the off-diagonal operators are frustration free, one only need to show that two configurations connected to each other by different sequences of moves are required to have the same relative weight in the superposition regardless of the path of counting weights. This kind of non-trivial cycles only appears for double plaquette moves that can be decomposed into two single plaquette moves. However, since the preferred ground state is the \emph{uniform} superposition of loop configurations, for both plaquette or double plaquette operator, this is never a problem. As one deforms the Hamiltonian to favor weighted superposition of loop configurations, or superposition of alternating signs as in the double semion model~\cite{PhysRevB.71.045110}, this becomes a non-trivial problem and could be an interesting question for future work.

\subsection{Ergodicity}

\begin{figure}[t!bh]
	\centering
	\subfigure[]{\label{fig:ergo1}\includegraphics[width=0.4\linewidth]{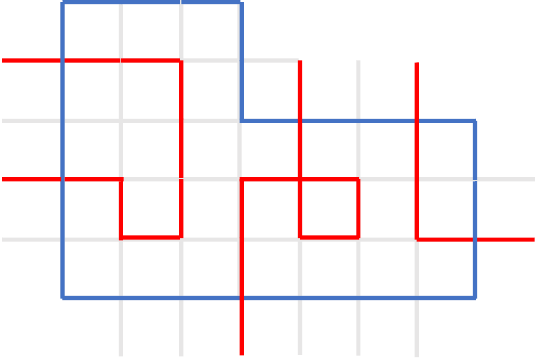} }
	\subfigure[]{\label{fig:ergo2}\includegraphics[width=0.4\linewidth]{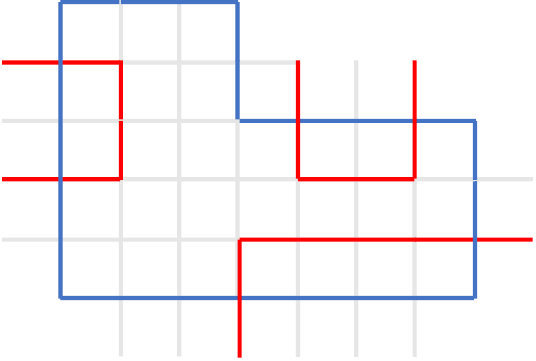} }
	
	\subfigure[]{\label{fig:ergo3}\includegraphics[width=0.4\linewidth]{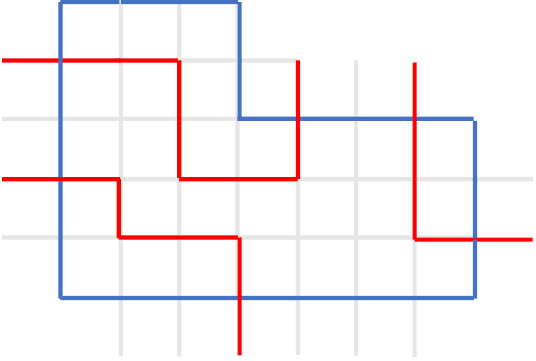} }
	\subfigure[]{\label{fig:ergo4}\includegraphics[width=0.4\linewidth]{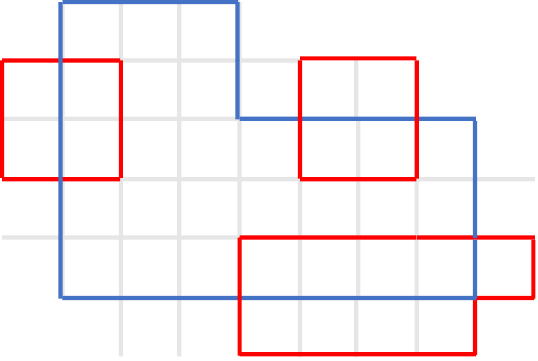} }
	
	\subfigure[]{\label{fig:ergo5}\includegraphics[width=0.4\linewidth]{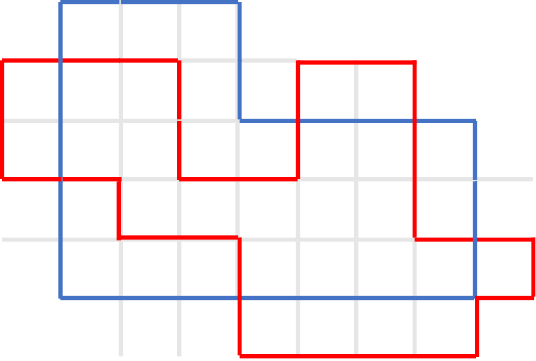} }
	\subfigure[]{\label{fig:ergo6}\includegraphics[width=0.4\linewidth]{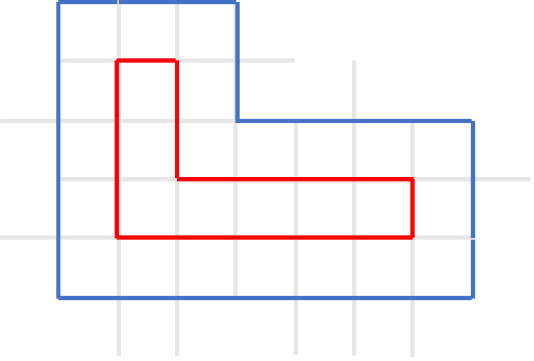} }
	\caption{Procedure to bring a generic bicolor loop configuration (a), to two choices of connectivity between the intersecting red loops staying next to the boundary blue loop inside (b) and (c). Combined with similar operations outside the blue loop, the red loops can either be directly removed using $C_{<f,f'>}$ operators from (d), or move completely inside the enclosing blue loop from (e), which can subsequently removed as in (f).}
	\label{fig:ergo}
\end{figure}

The general strategy if the ergodicity proof is to refer to the ergodicity of the toric code Hamiltonian whenever restricted to the subspace involving only two of the three states which removes the intersection of loops in the same color, and only resort to the $C_{<f,f'>}$ operators when necessary to remove intersection between loops in different colors. 

Starting from any generic bicolor loop configuration, first view the blue loops as never intersecting, meaning crossing will be interpreted as two loops touching at their corners. These blue loops partition the whole lattice into separate regions, within each of which there are only uncovered and red edges, as inside the blue boundary of Fig.~\ref{fig:ergo1}. Notice that this applies just as well to regions sandwiched between two nesting blue loops. Within each of these regions, the ergodicity of the toric code Hamiltonian implies that the repetitive action of the $A_f^{(1)}$ operators can bring all the red strings intersecting the blue boundary to stay next to the boundary between intersections, and remove all the closed red strings, resulting in either of the two outcomes depicted in Fig.~\ref{fig:ergo2} and ~\ref{fig:ergo3}. In combination with similar operations on the other side of the blue boundary, the red loops become either a set of disconnected closed loops as in Fig.~\ref{fig:ergo4} or a single closed loop winding around the blue boundary as in Fig.~\ref{fig:ergo5}. In the former case, each red loop can be shrunk two faces smaller at a time by the double plaquette operator, before eventually annihilated altogether. In the latter case, the double plaquette operators bring the red loop completely inside the blue boundary and can be shrunk and removed with single plaquette operator $A_j^{(1)}$. Either way, every red loop in the lattice disappears as a result. In the final step, the ergodicity of the $A_j^{(2)}$ operators are called to remove the blue loops. Since every loop configuration is connected to the empty configuration by the procedure above, we conclude that there is a unique ground state in the topological sector without non-contractible loops. The proof for the sectors involving non-contractible loops are completely analogous.

\subsection{Comparison with Reidemeister moves}

\begin{figure}[t!bh]
	\centering
	\includegraphics[width=0.8\linewidth]{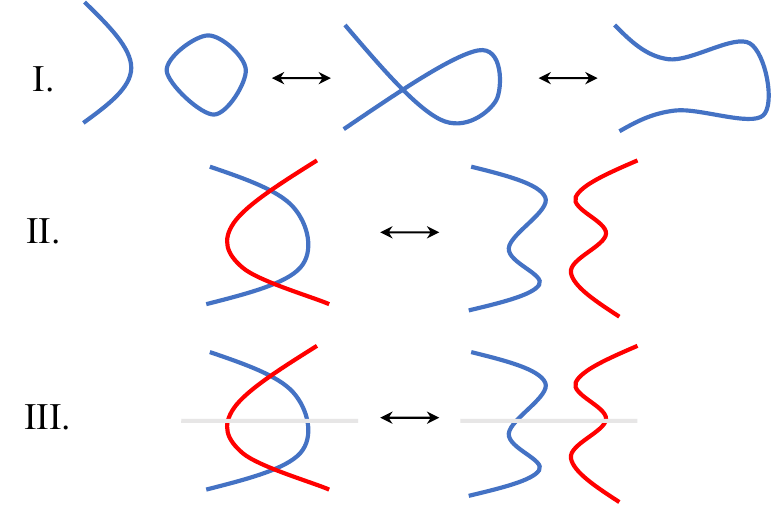}
	\caption{The three types of moves used in the proof of ergodicity. Type I and II are implemented by the single plaquette operator $A^{(a)}$'s, while type III is due to the double plaquette operator $C^{(a)}$'s.}
	\label{fig:Reidemeister}
\end{figure}

In the ergodicity proof, we have used three types of ``equivalence moves'', as shown in Fig.~\ref{fig:Reidemeister}, to relate classical configurations that appear in the same topological sector. The first two types of moves are similar to the Reidemeister moves of ambient isotopy, even though the latter distinguish knots from unknots where loops braid instead of intersecting with each other.

\section{Wilson loop operator algebra}
\label{sec:Wilson}

\begin{figure}[t!bh]
	\centering
	\includegraphics[width=0.6\linewidth]{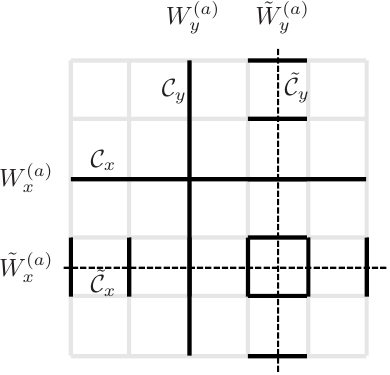}
	\caption{Definition of the Wilson loop operators, each operator in the figure comes in two different independent colors.}
	\label{fig:Wilson}
\end{figure}

Define the Wilson loop operators around non-contractible cycles $\mathcal{C}_{x,y}$ and on the dual lattice $\tilde{\mathcal{C}}_{x,y}$ as
\begin{equation}
	W_{x,y}^{(a)}=\prod_{j\in \mathcal{C}_{x,y}}X^{(a)}_j, \quad \text{and} \quad \tilde{W}_{x,y}^{(a)}=\prod_{j\in \tilde{\mathcal{C}}_{x,y}}Z^{(a)}_j,
\end{equation}
for $a=1,2,3$. They commute with each other except 
\begin{equation}
	\{W^{(a)}_{x},\tilde{W}^{(a)}_{y}\}=0, \quad \text{and } \{W^{(a)}_{y},\tilde{W}^{(a)}_{x}\}=0, 
\end{equation}for $a=1,2,3$, and
\begin{equation}
	\begin{aligned}
		\{W^{(1)}_{x,y},W^{(2)}_{x,y}\}=&W^{(3)}_{x,y},\\
		\{W^{(2)}_{x,y},W^{(3)}_{x,y}\}=&W^{(1)}_{x,y},\\
		\{W^{(3)}_{x,y},W^{(1)}_{x,y}\}=&W^{(2)}_{x,y},\\
		\{\tilde{W}^{(2)}_{x,y},W^{(1)}_{y,x}\}=&-W^{(1)}_{y,x},\\
		\{\tilde{W}^{(3)}_{x,y},W^{(1)}_{y,x}\}=&-W^{(1)}_{y,x},\\
		\{\tilde{W}^{(1)}_{x,y},W^{(2)}_{y,x}\}=&-W^{(2)}_{y,x},\\
		\{\tilde{W}^{(3)}_{x,y},W^{(2)}_{y,x}\}=&W^{(2)}_{y,x},\\
		\{\tilde{W}^{(1)}_{x,y},W^{(3)}_{y,x}\}=&W^{(3)}_{y,x},\\
		\{\tilde{W}^{(2)}_{x,y},W^{(3)}_{y,x}\}=&W^{(3)}_{y,x}.
	\end{aligned}
\end{equation}
Furthermore, they do not commute with the Hamiltonian due to the non-commutativity between operators with different superscripts acting on different subspaces of the local Hilbert space. 
Hence, acting the Wilson loop operators would necessarily bring the state out of the ground state manifold and the 16-fold degenerate ground states do not form an irreducible representation of the Wilson loop operator algebra.

\end{document}